\documentclass[letterpaper,10pt]{article}
\pdfoutput=1

\usepackage[margin=1in]{geometry}

\usepackage[utf8]{inputenc}
\usepackage[T1]{fontenc}
\usepackage[american]{babel}
\usepackage{textcomp}
\usepackage{csquotes}

\usepackage{amsmath}
\usepackage{amssymb}

\usepackage{booktabs,multicol}

\usepackage{color}
\usepackage[table]{xcolor}

\definecolor{lightgray}{gray}{0.9}
\definecolor{blueish}{rgb}{0.2,0.2,0.8}
\usepackage{graphicx}

\usepackage[utf8]{inputenc}
\usepackage[T1]{fontenc}
\usepackage[american]{babel}
\usepackage{textcomp}
\usepackage{csquotes}

\usepackage{authblk}

\usepackage[square,authoryear,sort]{natbib}
\usepackage[breaklinks=true,pdfborder={0 0 0},colorlinks=true]{hyperref}
\usepackage[all]{hypcap}
\usepackage{url}
\hypersetup{
    colorlinks,
    linkcolor={red!50!black},
    citecolor={blue!50!black},
    urlcolor={blue!80!black}
}

\title{The challenge and promise of software citation for credit, identification, discovery, and reuse}

\author{
Kyle E.\ Niemeyer\\
School of Mechanical, Industrial, and Manufacturing Engineering,\\
Oregon State University\\
\href{mailto:kyle.niemeyer@oregonstate.edu}{kyle.niemeyer@oregonstate.edu}
\and
Arfon M.\ Smith\\
GitHub Inc.\\
\href{mailto:arfon@github.com}{arfon@github.com}
\and
Daniel S.\ Katz\\
National Center for Supercomputing Applications \& \\ 
Electrical and Computer Engineering Department \& School of Information Sciences,\\
University of Illinois at Urbana--Champaign\\
\href{mailto:d.katz@ieee.org}{d.katz@ieee.org}
}

\date{}

\begin{document}

\maketitle

\section{Introduction}

Modern science and engineering research depends on software.
A 2009 survey of scientists found that 91\% consider software important or very important to their research~\citep{Hannay:2009wp}.
The scientific community uses citation to acknowledge traditional research results published in archival journals and conferences, but no such accepted standard exists to credit the considerable efforts that go into software---and most research software is not cited~\citep{Pan2015}.

One method to increase the amount of software developed, shared, and credited is to treat a software release as a publication.
There is good evidence that academics respond to incentives, including interview and survey data saying that increased citation would drive increased software development and sharing~\citep{howison-herbsleb2011,huang2013}.
Furthermore, evidence indicates that research activity increases when outputs can be formally counted~\citep{mcnaught2015}.
Finally, evidence of success in direct citations to datasets~\citep{belter2014}---as opposed to indirect citations through publications---suggests similar positive benefits for software citation.

In addition to the above benefits, since research results depend on the specific software used (e.g., version), proper citation---and the associated preservation---is necessary to ensure reproducibility~\citep{Sufi:2014en}.
Provenance of research results and data requires, among other things, a record of the software used to generate or process that data~\citep{Sandve:2013gh,Wilson:2014aa}.
Potential errors in software or variations due to environment~\citep{Morin:2012hz,Soergel:2015aa} further warrant the citation of the specific software used.

Different communities follow widely varying practices for citing software, with guidelines ranging from citations of an associated paper or the software itself via DOI~\citep{AAS:2016} to no policy at all, with many ad hoc practices in between~\citep{Howison2015}.
Furthermore, some research communities have not yet adopted open-source mentalities regarding research software.
Questions also remain about the role of curating\slash reviewing software---if software will be cited in the same manner as a publication, is quality assessment needed (e.g., peer code review)?
If so, how, when, and by whom?

\textbf{The main challenge of software citation is the lack of a way to uniquely identify released\slash published software, so those who use it can cite it.}
We also need a process to curate and review software.
As a first step towards sustainable, reusable, and attributable software, efforts are underway to establish citation practices for software used in research.

\section{Related Work}

Multiple organizations, including the WSSSPE Software Credit working group~\citep{WSSSPE1,WSSSPE2,WSSSPE3} and FORCE11 Software Citation Working Group~\citep{SoftwareCitation2016}, are working to standardize software citation practices.
Some of the observations and recommendations made here come from their efforts, which follow similar work by DataCite and the~\citet{DataCitation2014} to standardize research data citation practices.
More detail on these efforts, as well as broader related work by the community aimed at both citation and credit for software, can be found in the declaration of Software Citation Principles~\citep{SoftwareCitation2016}.

Even with an adopted, standard method for software citation, indexers (e.g., Web of Science, Scopus, Google Scholar) currently lack support for indexing such citations.
The astrophysics community represents a possible exception, where the community-run service NASA ADS~\citep{nasaads} also carries out indexing of citations.
The ADS already indexes references to software listed in the \citet{ascl}, and has made a commitment to expand upon this functionality following an AAS\slash GitHub-sponsored meeting~\citep{aas-software-index}.

Finally, peer review of software distinctly lacks a standard practice.
Some journals that accept both ``software'' submissions and general research papers can conflate review of the submitted software with a review of research output \textbf{produced by} the software.
Software-only journals (e.g., \textit{Journal of Open Research Software}, \textit{SoftwareX}) and the rOpenSci community represent exceptions to this, with documented review processes.
For example, rOpenSci has a process \citep{rOpenSci-reviews} used to determine whether a package can become part of the rOpenSci collection.

\section{Challenges and Research Directions}

Table~\ref{tab:challenges} summarizes key challenges and research directions for software citation, along with possible solutions\slash methods.
\rowcolors{2}{}{lightgray}
\begin{table}[htbp]%
\centering
\caption{Key research challenges and possible solutions\slash methods\label{tab:challenges}}{%
\begin{tabular}{@{}p{0.37\linewidth}p{0.60\linewidth}@{}}
\toprule
\textbf{Key research challenge} & \textbf{Possible solutions/methods} \\
\midrule

Identify necessary metadata associated with software for citation &
We suggest metadata below, and the CodeMeta project~\citep{CodeMeta} is working to determine minimal metadata. \\

Standardize proper formats for citing software in publications &
The FORCE11 Software Citation Working Group is defining, and gaining community acceptance for, software citation principles, after which a follow-on group will begin implementation efforts. \\

Establish mechanisms for software to cite other software (i.e., dependencies) &
Software publication as software papers allows this. For software that is directly published, this is an open challenge.\\ 

Develop infrastructure to support indexing of software citations 
within the existing publication citation ecosystem &
The FORCE11 Software Citation Implementation Working Group will also work on this, in collaboration with publishers and indexers. \\

Determine standard practices for peer review of software &
Professional societies and science communities need to determine how this will happen. \\

Increase cultural acceptance of the concept of software as a digital product &
Acceptance will happen over time; unclear how to accelerate this process. \\
\bottomrule
\end{tabular}}
\end{table}

For software to be cited, we recommend that metadata include software name, primary authors\slash contributors (name and ORCID), DOI or other unique and persistent identifier, and location where the software has been published\slash archived.
Typically, DOI (or a similar unique identifier) will provide both identification and location, but if not then metadata should include a way to locate the software (e.g., URL).
This information should be provided in a \texttt{CITATION} file, potentially in JSON or XML format (and with an appropriate metadata schema, e.g., DOAP) to allow automatic processing.
Citations of software in publications should minimally include software name, primary authors, and DOI or location where the software was published\slash archived; however, individual citation formats will vary based on the particular style of journals, conferences, or professional societies.
Existing services such as Libraries.io~\citep{Libraries.io} and Depsy~\citep{Depsy} automate software dependency tracking; these could be harnessed to produce citation networks.
Technical solutions to these challenges exist---community acceptance is instead needed.

In arguing the need for software citation, we simultaneously introduce the question of \textbf{when} software should be cited.
Although there is no clear answer to this question, ensuring reproducibility of research results requires citation of software if used directly and important to research results.
In other words, if using different software could produce different data or results, then the software used should be cited.

Research and development efforts are needed to solve the remaining challenges.
These include: how can we cite closed-source\slash commercial software---can the above information be provided, even if the software itself is not publicly preserved?
Software citations require indexing to create a citation network akin to publications to carry weight for academic credit and reputation, so how can index services be encouraged to fully index software?
How can publications indicate direct use of software for research in citations, where results would not be possible without efforts of software authors---should such a citation be ``weighted'' higher than others?
The \citet{AAS:2016} suggests a new ``Software'' section below the acknowledgements; other approaches include machine-readable supplementary data~\citep{Katz:2015aa}.

Finally, open questions remain on whether citable software should go through peer review and, if so, how can this be implemented?
Should citable software itself follow the arXiv preprint model where releases are made available for users and the community to judge, or, alternatively, the software paper model where ``advertising'' papers undergo peer review in a relevant community?

\section*{Acknowledgments} \label{sec:acks}
Some work by K.\ E.\ Niemeyer was supported by NSF grant ACI-1535065.
Work by D.\ S.\ Katz was supported by the National Science Foundation (NSF) while working at the Foundation; any opinion, finding, and conclusions or recommendations expressed in this material are those of the author and do not necessarily reflect the views of the NSF.
The authors wish to thank the members of the SciSIP community and mailing list, in particular Chris Belter, James Howison, and Joshua Rosenbloom, for a useful discussion on incentives for software publications and how their impact has been and could further be measured.
In addition, the authors acknowledge members of the FORCE11 Software Citation Working Group, whose views contributed to this document.

\bibliographystyle{ACM-Reference-Format-Journals}
\bibliography{SC-challenge}

\end{document}